\documentclass[twocolumn]{aastex631}
\usepackage{graphicx}
\usepackage{amsmath}
\usepackage{multirow,soul}

\shorttitle{Mercury's chaotic secular evolution as a subdiffusive process}
\shortauthors{Abbot et al.}
\begin{document}

\title{Mercury's chaotic secular evolution as a subdiffusive process}

\correspondingauthor{Dorian S. Abbot}
\email{abbot@uchicago.edu}

\author{Dorian S. Abbot}
\affiliation{Co-first author \\}
\affiliation{Department of the Geophysical Sciences \\
The University of Chicago \\
Chicago, IL 60637 USA}

\author{Robert J. Webber}
\affiliation{Co-first author \\}
\affiliation{Department of Computing \& Mathematical Sciences \\
California Institute of Technology \\
Pasadena, California 91125 USA}

\author{David M. Hernandez}
\affiliation{Department of Astronomy \\
Yale University \\
New Haven, CT 06511 USA}

\author{Sam Hadden}
\affiliation{Canadian Institute for Theoretical Astrophysics \\
University of Toronto \\
Toronto, ON M5S 3H8, Canada}

\author{Jonathan Weare}
\affiliation{Courant Institute of Mathematical Sciences \\ 
New York University \\ 
New York, NY 10012 USA}

%%%%%%%%%%%%%%%%%%%%%%%%%%%%%%%%%%%%%%%%%%%%%%%%%%%%%%%%%%%%%%%%%%%%
\begin{abstract}
Mercury's orbit can destabilize, generally resulting in a collision with either Venus or the Sun. Chaotic evolution can cause $g_1$ to decrease to the approximately constant value of $g_5$ and create a resonance. Previous work has approximated the variation in $g_1$ as stochastic diffusion, which leads to a phenomological model that can reproduce the Mercury instability statistics of secular and $N$-body models on timescales longer than 10~Gyr. Here we show that the diffusive model significantly underpredicts the Mercury instability probability on timescales less than 5~Gyr, the remaining lifespan of the Solar System. This is because $g_1$ exhibits larger variations on short timescales than the diffusive model would suggest. To better model the variations on short timescales, we build a new subdiffusive phenomological model for $g_1$. Subdiffusion is similar to diffusion but exhibits larger displacements on short timescales and smaller displacements on long timescales. We choose model parameters based on the behavior of the $g_1$ trajectories in the $N$-body simulations, leading to a tuned model that can reproduce Mercury instability statistics from 1--40~Gyr. This work motivates fundamental questions in Solar System dynamics:  Why does subdiffusion better approximate the variation in $g_1$ than standard diffusion? Why is there an upper bound on $g_1$, but not a lower bound that would prevent it from reaching $g_5$?
\end{abstract}

%%%%%%%%%%%%%%%%%%%%%%%%%%%%%%%%%%%%%%%%%%%%%%%%%%%%%%%%%%%%%%%%%%%%
\section{Introduction} \label{sec:intro}

Since the landmark study of \citet{laskar1994large}, the potential for Mercury's orbit to destabilize has been widely recognized. The destabilization process has been studied both with simplified test particle secular models \citep{LithwickWu2011secular_chaos,Boue2012,lithwick2014secular,batygin2015chaotic}  and sophisticated, high-order secular models \citep{laskar2008chaotic,MogaveroLaskar2021,mogavero2022origin,hoang2022long,mogavero2023timescales},
as well as more computationally intensive and physically realistic $N$-body codes \citep{laskar2008chaotic,batygin2008dynamical,laskar2009existence,zeebe2015highly,zeebe2015dynamic,brown2020repository,brown2022long,brown2023general,abbot2021rare,abbot2023simple,hernandez2022stepsize}. 
The secular models have led to the insight that Mercury's orbit destabilizes due to resonance between the Solar System's $g_1$ and $g_5$ secular eigenfrequencies, which are primarily associated with Mercury and Jupiter, respectively. 

The inherent unpredictability of chaotic dynamical systems like the Solar System makes it necessary to describe the long-term evolution statistically. Statistical descriptions of how the phase space distribution of an ensemble of systems evolves over time are relatively well-developed for simple area-preserving planar maps and 2 degree-of-freedom systems \citep[e.g.,][]{Mackay1984,meiss_symplectic_1992,zaslavsky_chaos_2002}, though even in this simplest case there remain important unresolved questions \citep[e.g.,][]{meiss_thirty_2015}. Given the lack of theoretical understanding of chaotic transport in systems with a larger number of degrees of freedom, phenomenological models such as those discussed in this paper can serve as useful tools for describing complex real-world systems and can potentially provide clues for better understanding the underlying chaotic dynamics.

Several diffusive phenomenological models have been used to approximate the Solar System dynamics and predict the probability of Mercury instability events as a function of time. \citet{woillez2020instantons} analyzed the simplified secular Hamiltonian of \citet{batygin2015chaotic}, which considers Mercury to be massless and approximates the other planets as quasiperiodic, and they identified the slowly varying component of this Hamiltonian as driving Mercury's dynamics. They approximated the dynamics as diffusive with constant diffusivity, a reflecting upper boundary, and an absorbing lower boundary that signifies Mercury instability events. 

\begin{figure}[ht!]
\centering
\includegraphics[width=\linewidth]{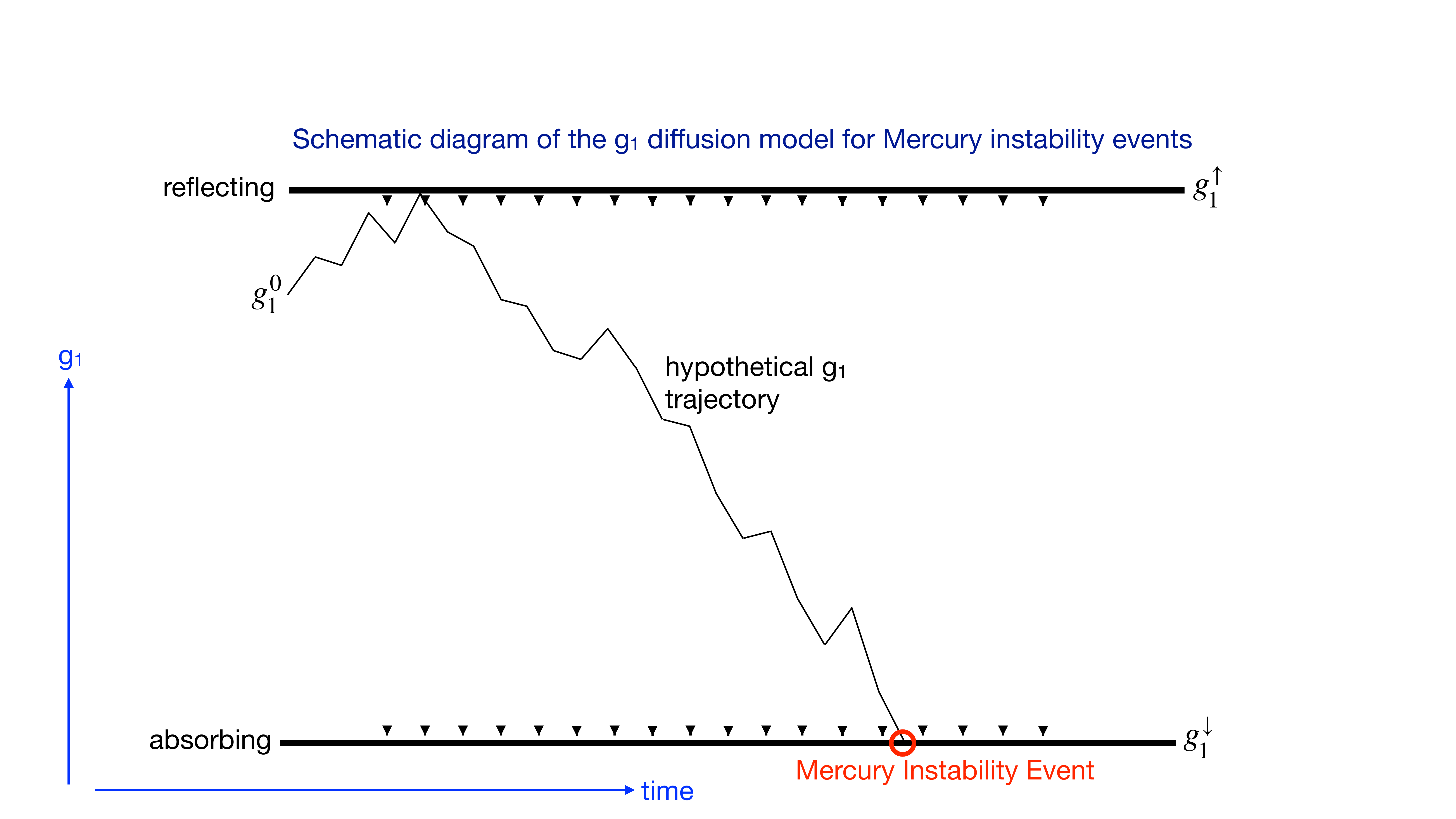}
\caption{Schematic diagram of the $g_1$ diffusion model for Mercury instability events. $g_1$ is initialized at $g_1^0$ and diffuses. $g_1^\uparrow$ is a hard, reflecting upper boundary and $g_1^\downarrow$ is an absorbing lower boundary. If $g_1$ reaches $g_1^\downarrow$, a Mercury instability event occurs.}
\label{fig:g1-diffusion}
\end{figure}

Later, \citet{MogaveroLaskar2021} speculated that the diffusive model might apply to the long-term variation of $g_1$ itself (see section~\ref{sec:g1_model} for the definition of $g_i$ as the Solar System's secular eigenfrequencies). They applied the diffusive model using $g_5$, which is effectively constant \citep{hoang2021chaotic}, as the absorbing lower boundary at which Mercury instability  occurs (Fig.~\ref{fig:g1-diffusion}). They tuned the upper boundary and diffusivity to produce a reasonable approximation to the Mercury instability probability statistics of a secular model on timescales longer than 10~Gyr, when at least 4\% of the simulations have gone unstable. The 10~Gyr timescale is longer than the future lifespan of the Sun, so their model can only be interpreted as an abstract investigation of the Solar System dynamics.

Recently, \citet{brown2023general} compared the $g_1$ diffusive model \citep{MogaveroLaskar2021} to the 5~Gyr $N$-body simulations performed by \citet{laskar2009existence}. They found what appeared to be reasonable correspondence; however, in their Figures 2--4, they plotted one minus the probability of a Mercury instability event, which obscures the difference between small probabilities spanning orders of magnitude. \citet{brown2023general} also compared the $g_1$ diffusive model to their own $N$-body simulations with general relativity artificially either fully or partially disabled. Similar to the $N$-body simulations in this paper, they approximated general relativity as a simple potential. Their plots suggest that the diffusive model provides a qualitatively reasonable approximation of the evolution of the Mercury instability probability when it is above $\approx$5\%, but the plots obscure smaller probabilities.  In summary, the $g_1$ diffusive model (Fig.~\ref{fig:g1-diffusion}) can be tuned to approximate the Mercury instability probability produced by more complex secular or $N$-body models, as long as the Mercury instability probability exceeds $\approx$5\%.

In this paper we will show that the $g_1$ diffusive model significantly underpredicts the Mercury instability probability over the next 5~Gyr period in which the Sun will remain on the main sequence. We find that the discrepancy primarily results from the diffusive model producing variations of $g_1$ that are too small on timescales less than $\sim$0.3~Gyr. To better model the short-time variations of the $g_1$ trajectories, we propose a $g_1$ subdiffusive model \cite[see][for an introduction to subdiffusion]{henry2010introduction}. 
The subdiffusive model is consistent with the work of \citet{hoang2021chaotic}, who found that $g_1$ subdiffuses by fitting a power law to the standard deviation of $g_1$ in a large ensemble of secular models as a function of time. We fit the parameters of our subdiffusive model to the mean square displacement of $g_1$, the probability density function (pdf) of $g_1$ from $N$-body simulations,
and the fraction of $N$-body simulations that reach the instability threshold.
This produces a five-parameter stochastic subdiffusive model that closely approximates $N$-body Mercury instability probabilities as small as $\sim 10^{-4}$ (the lowest value we are able to estimate with $N$-body results due to limited sample size) and as high as $\sim 0.5$ (the highest value we are able to estimate with $N$-body results due to limited run lengths). The $g_1$ subdiffusive model is successful at reproducing a variety of statistics from the $N$-body code, suggesting that despite its simplicity, the model captures important aspects of the relevant dynamics.

%%%%%%%%%%%%%%%%%%%%%%%%%%%%%%%%%%%%%%%%%%%%%%%%%%%%%%%%%%%%%%%%%%%%
\section{Models} \label{sec:model}

\subsection{Model used to calculate \texorpdfstring{$g_1$}{g1} statistics}
\label{sec:g1_model}

We calculate $g_1$ statistics from the 2750-member \texttt{Fix dt} ensemble of simulations produced by \citet{abbot2023simple}. The simulations contain all 8 Solar System planets and no moons, asteroids, or comets.
The simulations use the \texttt{WHFAST} integration scheme \citep{rein2015whfast} from the REBOUND $N$-body code \citep{rein2012rebound}, which is a Wisdom-Holman scheme (WH) \citep{WH1991}.
The only parameterized physics scheme is an approximation of general relativity with a modified position-dependent potential \citep{nobili1986simulation}, which is implemented as the \texttt{gr\_potential} scheme in REBOUNDx \citep{Tamayo2020reboundx}.
We initialized the simulations with Solar System conditions on February 10, 2018 taken from the NASA Horizons database and then added a uniform grid of perturbations to Mercury's $x$-position, each separated by 10~cm.
We used a fixed time step of $\sqrt{10}\approx 3.16$~days, which we demonstrated was sufficiently small to produce converged Mercury instability statistics \citep{abbot2023simple}, and ran the simulations for 5~Gyr.
The trajectories of eccentricity and $g_1$ are nearly identical among the simulations over the first 138~Myr of the simulation and then begin to noticeably separate from each other on the order of $\approx$1\%. This is longer than the typical quoted value of $\sim$50~Myr for Solar System orbital calculations to diverge \citep[e.g.,][]{laskar2011strong,zeebe2017numerical}, possibly because Mercury's eccentricity and $g_1$ take longer to diverge than other variables.

The $g_i$ frequencies are defined through the following eigenfunction expansion \citep[Ch.~7]{murray1999solar}: 
\begin{align}
\label{eq:linear_1}
e_i \cos \varpi_i &= \sum_k M_{ik} \cos \alpha_k, \\
\label{eq:linear_2}
e_i \sin \varpi_i &= \sum_k M_{ik} \sin \alpha_k,
\end{align}
where the index $i$ ranges over the planets $i = 1, 2, \ldots, 8$,
$e_i$ denotes the eccentricity of each planet, and
$\varpi_i$ denotes the longitude of perihelion.
The values
$(M_{i,k})_{1 \leq i,k \leq 8}$ are the coefficients in the eigenfunction expansion, and the terms $\alpha_k = g_k t + \beta_k$ describe the angles of the oscillations.

The first-order approximation with constant parameter values \eqref{eq:linear_1}-\eqref{eq:linear_2} is accurate over the shortest period of oscillation $\min_i 2\pi / g_i$, but degrades thereafter.
Indeed, when we fit the approximation to the output of a more complex $N$-body simulation, the values $M_{ik}$, $g_k$, and $\beta_k$ are all slowly changing as a function of time.
Despite the imprecision, the first-order approximation leads to qualitative insights, as it correctly indicates the possibility for destabilization when two of the $g_i$ frequencies approach resonance.

To calculate $g_1$, we use the \texttt{frequency\_modified\_fourier\_transform} routine \citep{vsidlichovsky1997frequency} from the celmech package \citep{hadden2022celmech}, which requires a number of output times that is a power of 2. 
We find that $2^{10}$ output times are sufficient to obtain stable $g_i$ estimates from data, so we divide the \texttt{Fix dt} ensemble data
into blocks of 10.24~Myr, with each block containing $2^{10}$ output times. Our calculated values of $g_1$ are therefore the average value over each 10.24~Myr segment. We apply \texttt{frequency\_modified\_fourier\_transform} to $e_1 \exp(i \varpi_1)$ to calculate the four Fourier modes  with the highest amplitude for a given 10.24~Myr segment and chose $g_1$ to be the $g_i$ closest to $g_1$ from the previous segment.

\subsection{Model used to calculate 40~Gyr instability statistics}

As a new contribution of this paper, we perform 1000 extensions of the \texttt{Fix dt} simulations and make them publicly available at \texttt{https://archive.org/download/LongRun}. These extensions have exactly the same parameters as the \texttt{Fix dt} simulations described above, but we run them for a total of 40~Gyr.
As in \citet{abbot2021rare,abbot2023simple}, we define a Mercury instability event as occurring when Mercury passes within 0.01~AU of Venus and stop the simulations at that point. 
In \citet{abbot2023simple}, we showed that after $10^{12}$ time steps (10~Gyrs), roundoff relative error is of order $10^{-5}$ for the semimajor axis and order $10^{-9}$ for the energy. Roundoff error is growing as $\sim t^{0.5}$ at 10~Gyr, so it should remain small for the 40~Gyr simulations we performed.

\subsection{Ensemble of ensembles used to compute 5~Gyr instability statistics}

To compute Mercury instability statistics on a timescale of less than 5~Gyr, we use the ensemble of $N$-body ensembles constructed by \citet{abbot2023simple}, which includes the 2501-member \citet{laskar2009existence} ensemble, the 1600-member \citet{zeebe2015highly} ensemble, as well as both the 2750-member \texttt{Var dt} and 2750-member \texttt{Fix dt} ensembles of \citet{abbot2023simple}, for a total of 9601 members.

\subsection{Diffusive and subdiffusive models}
\label{sec:diff_subdiff_model}

Both the diffusive and subdiffusive models are defined using fractional Brownian motion.
Fractional Brownian motion is a mean-zero Gaussian process which we denote by $W(t)$ at each time $t \geq 0$.
It is defined as the unique mean-zero Gaussian process which starts from $W(0) = 0$ and has increments satisfying
\begin{equation}
\label{eq:definition}
    \langle |W(t) - W(s)|^2 \rangle = |t - s|^{2\alpha}
\end{equation}
at all times $t, s$. Here, $\alpha \in (0, 1)$ is the Hurst parameter, which is $\alpha = 1/2$ for a standard diffusion, whereas $\alpha \in (0, 1/2)$ for a subdiffusion and $\alpha \in (1/2, 1)$ for a superdiffusion \citep[Sec.~1.2]{henry2010introduction}. As a result of the scaling relation \eqref{eq:definition}, a subdiffusion exhibits larger variations over short timescales than a standard diffusion.

The simple diffusion model of \citet[Sec.~8.2]{MogaveroLaskar2021} states that the $g_1$ value at time $t$ (units of Gyrs) satisfies:
\begin{equation}
\label{eq:simplest}
    g_1(t) = 
    \begin{cases}
        g_1^0 + r W(t), & t \leq T \\
        g_1^\uparrow + r [W(t) - \max_{0 \leq s \leq t} W(s)], & t > T,
    \end{cases}
\end{equation}
where $W(t)$ is a standard Brownian motion, $r > 0$ is a scaling factor, 
$g_1(0) = g_1^0$ is the initial condition,
$g_1^\uparrow > g_1^0$ is a reflecting upper boundary condition,
and
\begin{equation}
    T = \min\{t \geq 0: g_1^0 + rW(t) \geq g_1^\uparrow\}
\end{equation}
is the first time the process hits the upper boundary.
The process advances forward until hitting the lower boundary $g_1^\downarrow$ at a random time
\begin{equation}
\label{eq:stopping}
    \tau = \min\{t \geq 0: g_1(t) = g_1^\downarrow\}.
\end{equation}
Then, instability occurs in the model (Fig.~\ref{fig:g1-diffusion}).

\citet[Sec.~8.2]{MogaveroLaskar2021} motivate their use of diffusion and the reflecting upper boundary by observing that the pdf of $g_1$ has Gaussian tail behavior at low $g_1$ values, but drops off sharply at high $g_1$ values \citep{hoang2021chaotic}.
\cite{woillez2017long} provide additional motivation, by referencing the theory of slow-fast dynamical systems \citep[Ch.~6]{gardiner2009stochastic}, in which the evolution of a slow variable can sometimes be modeled as a diffusive SDE.
However, \cite{hoang2021chaotic} argue that the time evolution of $g_1$ under the secular equations matches a subdiffusion more closely than a standard diffusion.

Next, we propose a more general model in which we allow $W(t)$ to be a fractional Brownian motion with a Hurst parameter $\alpha$ not necessarily equal to $1/2$. In the general model, the $g_1$ value at time $t$ is given by the same equations \eqref{eq:simplest}-\eqref{eq:stopping} but the standard Brownian motion is replaced by a fractional Brownian motion. As before, the process advances forward until hitting the lower boundary $g_1^\downarrow$. Then, instability occurs.

To simulate from the diffusive and subdiffusive models, we do not invoke equation \eqref{eq:simplest} directly, because this leads to a discretization error of size $\mathcal{O}(\delta^\alpha)$ for a time step $\delta$ \citep{7822095}.
Therefore, we pursue an alternative, more efficient discretization strategy \citep{PhysRevE.97.020102}. First, we generate a random vector containing the values of the fractional Brownian motion at discrete output times $\mathbf{W} = \bigl(W(\delta), W(2 \delta), \ldots, W(N \delta)\bigr)$, using the algorithm of \cite{DN97} as implemented in the \textsf{stochastic} package for python \citep{Fly222}.
Next, we set $g_1(t=0) = g_1^0$ and apply the recursive update formula
\begin{equation}
\label{eq:approximate}
\begin{aligned}
    \tilde{g}_1(t) &= g_1(t - \delta) + r \bigl[W(t) - W(t - \delta)\bigr] \\
    g_1(t) &= g_1^{\uparrow} - |g_1^{\uparrow} - \tilde{g}_1(t)|.
\end{aligned}
\end{equation}
This formula directly invokes a fractional Brownian motion that is reflected off an upper boundary.
The formula yields the exact correct distribution without any discretization error for $\alpha = 1/2$ \cite[pg.~393]{doob1953} and closely approximates the distribution for all $\alpha \in (0, 1)$. We apply equation \eqref{eq:approximate} with a time step $\delta = 0.01$~Gyr in our simulations, and our code is available at \texttt{https://knowledge.uchicago.edu/record/10351}.

\begin{table}[t]
\begin{center}
\begin{tabular}{ l|l|l|l } 
\textbf{parameters} & \textbf{ML21} &
\textbf{diffusive} & \textbf{subdiffusive} \\
 \hline
 $g_1^0$ ($''\,$yr$^{-1}$) & $5.58$ & $5.60$  & $5.60$   \\ 
 $g_1^\uparrow$ ($''\,$yr$^{-1}$) & $5.72$ & $5.79$  & $5.85$  \\ 
 $g_1^\downarrow$ ($''\,$yr$^{-1}$) & $4.26$ & $4.55$  & $4.86$  \\
 $\alpha$ & $0.50$ & $0.50$  & $0.26$  \\
 $r$ ($''\,$yr$^{-1}$ Gyr$^{-\alpha}$) & $0.20$ & $0.17$  & $0.16$  
\end{tabular}
\end{center}
\caption{Parameters and their values for the diffusive and subdiffusive models. The parameter values for the ML21 diffusive model are taken from \citet{MogaveroLaskar2021}, but rewritten using our variables.}
\label{tab:params}
\end{table}

%%%%%%%%%%%%%%%%%%%%%%%%%%%%%%%%%%%%%%%%%%%%%%%%%%%%%%%%%%%%%%%%%%%%
\section{Results} \label{sec:results}

\subsection{Problems with the \texorpdfstring{$g_1$}{g1} diffusive model} \label{sec:diffusive}

\begin{figure}[t]
\centering
\includegraphics[width=\linewidth]{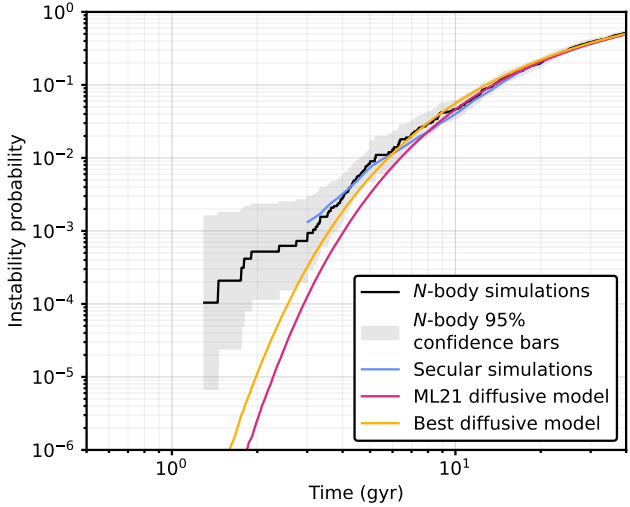}
\caption{Probability that Mercury's orbit becomes unstable as a function of time for the $N$-body simulations (black),
the \citet{MogaveroLaskar2021} secular simulations (blue),
the \citet{MogaveroLaskar2021} diffusive model (magenta),
and the best fit diffusive model (yellow). We obtained the \citet{MogaveroLaskar2021} secular simulation results using a data digitizer.}
\label{fig:DiffHits}
\end{figure}

In this subsection, we point out several issues with the $g_1$ diffusive model. As a starting point, the $g_1$ diffusive model is not effective at predicting Mercury instability probabilities over physically realistic timescales.
On timescales longer than $\sim$10~Gyr, the model matches with the instability probabilities from secular model simulations \citep{MogaveroLaskar2021} and from our 40~Gyr $N$-body simulations (Fig.~\ref{fig:DiffHits}). However, on the 5~Gyr timescale of the future of the Solar System, the diffusive model underpredicts Mercury instability events by a factor of 3--1000 (Fig.~\ref{fig:DiffHits}).

\begin{figure}[t]
\centering
\includegraphics[width=\linewidth]{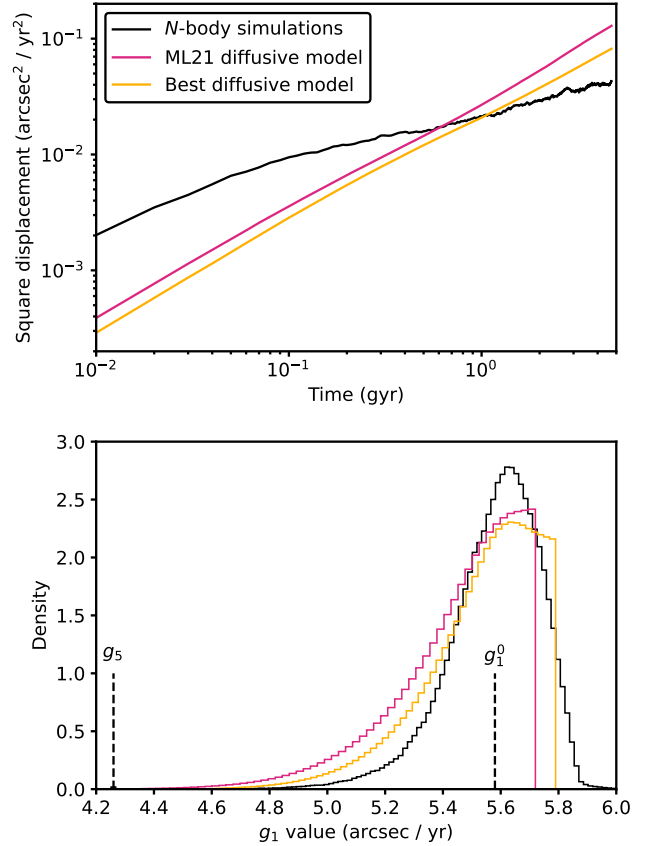}
\caption{Top: Mean square displacement of $g_1$ as function of time offset $\Delta t$ for the $N$-body model (black), the \cite{MogaveroLaskar2021} diffusive model (magenta), and the best fit diffusive model (yellow). 
Bottom: Probability density function (pdf) of $g_1$ from the $N$-body model (black) and the two diffusive models (magenta and yellow). The pdfs include all $g_1$ data over 5~Gyr from the 2750-member $N$-body ensemble and an equivalent generated diffusive ensemble. The values of $g_5$ and $g_1^0$ for the \cite{MogaveroLaskar2021} diffusive model are indicated on the plot.}
\label{fig:DiffLocal}
\end{figure}

One possible explanation for the fact that the $g_1$ diffusion produces too few Mercury instability events on shorter timescales ($<$10~Gyr) is that subdiffusion \citep{henry2010introduction} better approximates the chaotic evolution of $g_1$. The main characteristic of subdiffusion is that it exhibits larger displacements on short timescales and smaller displacements on long timescales than diffusion. To investigate this idea quantitatively, we calculate the mean square displacement $\langle |\Delta g_1|^2 \rangle$ across a range of time offsets $\Delta t$ and consider a scaling relationship
\begin{equation}
    \langle |\Delta g_1|^2 \rangle \sim |\Delta t|^{2\alpha}
    \label{eq:MeanSquare}
\end{equation}
where $\alpha \in (0, 1)$ is the Hurst parameter that can be chosen to match the data. As explained in section~\ref{sec:diff_subdiff_model}, $\alpha = 1/2$ corresponds to standard diffusion, whereas $\alpha \in (0, 1/2)$ corresponds to subdiffusion and $\alpha \in (1/2, 1)$ corresponds to superdiffusion. 
The top panel of Fig.~\ref{fig:DiffLocal} shows that $\alpha$ for the $N$-body model is much less than $1/2$ (we will show below that $\alpha \approx 0.26$), and the diffusive model produces a mean square displacement for $g_1$ that is too small on timescales less than $\sim$0.3~Gyr.

Next, let us compare the pdf of $g_1$ between the $N$-body model and the diffusive model (Fig.~\ref{fig:DiffLocal} bottom). The diffusive model as tuned by \citet{MogaveroLaskar2021} overpredicts by an order of magnitude the probability that $g_1$ has a value less than 5$\,''\,$yr$^{-1}$. The unrealistic low values of $g_1$ occur in the diffusion model because the lower boundary on $g_1$ is set to be $g_5 = 4.257\,''\,$yr$^{-1}$. However, while the main physical mechanism for a Mercury instability event is a $g_1$-$g_5$ resonance \citep{batygin2015chaotic}, the $g_5$ resonance might cause non-diffusive behavior as $g_5$ is approached. For example, the $g_1$ trajectories that lead to Mercury instability events in Figure 5 of \citet{MogaveroLaskar2021} often show large, erratic jumps from a value of $\sim5\,''\,$yr$^{-1}$ to $g_5$ as the instability event occurs. When we allow the lower boundary on $g_1$ to be a free parameter in the best-fit diffusive model, the fit for small $g_1$ improves by about a factor of two, but a large discrepancy remains in order for the model to match the instability probability.

To conclude this subsection, the $g_1$ diffusive model has the following defects:
\begin{enumerate}
\item It underpredicts the Mercury instability probability on timescales less than 10~Gyr.
\item It produces too small variations in $g_1$ on timescales less than $\sim$0.3~Gyr. 
\item It leads to a scaling of the mean square displacement $\langle |\Delta g_1|^2 \rangle$ with the time offset $\Delta t$ that does not fit the $N$-body simulations.
\item As formulated by \citet{MogaveroLaskar2021}, it assumes that $g_1$ must diffuse all the way to $g_5$ to produce an instability event, which is not the case.
\end{enumerate}

\subsection{An improved \texorpdfstring{$g_1$}{g1} subdiffusive model} \label{sec:subdiffusive}

\begin{figure}[t]
\centering
\includegraphics[width=\linewidth]{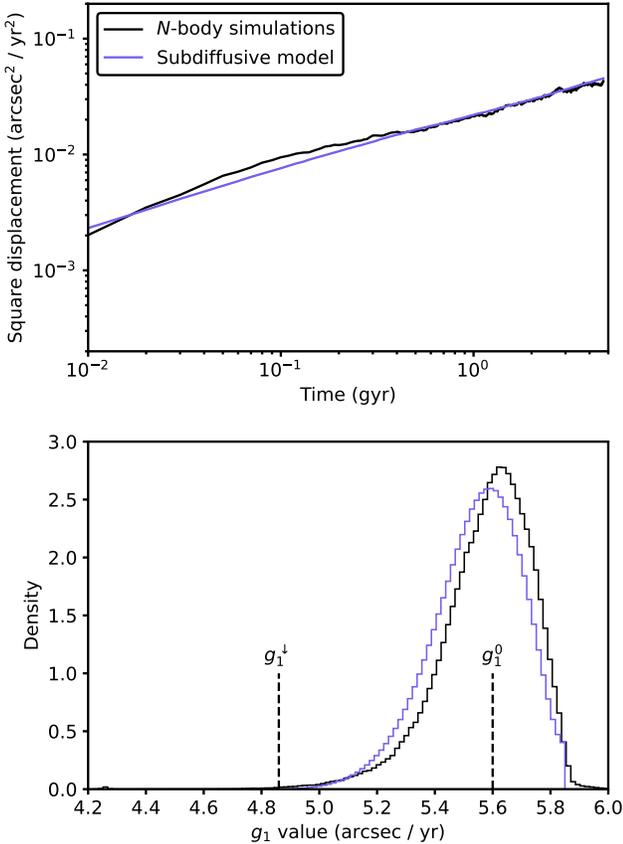}
\caption{Top: Mean square displacement of $g_1$ as function of time offset ($\Delta t$) for the $N$-body model (black) and the subdiffusive model described in Secs.~\ref{sec:subdiffusive}-\ref{sec:tuning} (purple). 
Bottom: Probability density function of $g_1$ from the $N$-body model (black) and the subdiffusive model (purple). The pdfs include all $g_1$ data over 5~Gyr from the 2750-member $N$-body ensemble and an equivalent generated subdiffusive ensemble. The values of $g_5$ and $g_1^0$ are indicated on the plot.}
\label{fig:SubDiffLocal}
\end{figure}

We now apply our new subdiffusive model that addresses the limitations of the $g_1$ diffusive model. 
To begin, we fix the initial value $g_1^0 = 5.60\,''\, \text{yr}^{-1}$, because this is the $g_1$ value at which the $N$-body simulations start to diverge.
Then, we make the modeling assumption that the mean square displacement scales as a power law
\begin{equation}
\label{eq:new_scaling}
\langle |\Delta g_1|^2 \rangle = r^2 |\Delta t|^{2\alpha}
\end{equation}
over short timescales $\Delta t$.
This modeling assumption fits the mean square displacement data for the $N$-body simulations with $\alpha=0.26$ and $r = 0.16\,''\,\text{yr}^{-1}$ (see Fig.~\ref{fig:SubDiffLocal} top). 
In previous work, \citet{hoang2021chaotic} estimated the same value $\alpha=0.26$ by fitting a power law to the standard deviation of $g_1$ in a large ensemble of secular models as a function of time.

\begin{figure}[h]
\centering
\includegraphics[width=\linewidth]{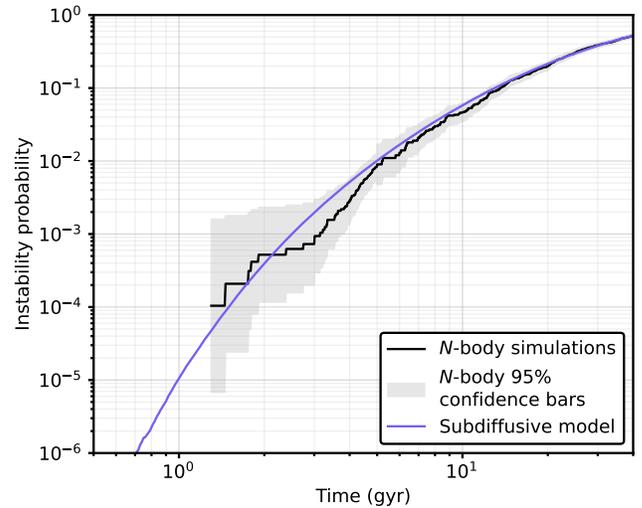}
\caption{Probability that Mercury's orbit becomes unstable as a function of time for the $N$-body model (black) and the subdiffusive model described in Secs.~\ref{sec:subdiffusive}-\ref{sec:tuning} (purple). This figure should be compared with Fig.~(\ref{fig:DiffHits}), which gives the equivalent information for the \citet{MogaveroLaskar2021} diffusive model.}
\label{fig:SubDiffHits}
\end{figure}

Next, we impose an absorbing lower boundary $g_1^\downarrow$ and a reflecting upper boundary $g_1^\uparrow$. These boundary conditions fit the probability density function (pdf) of $g_1$ from the $N$-body simulations for $g_1^\downarrow = 4.85\,''\, \text{yr}^{-1}$, $g_1^\uparrow = 5.79\,''\, \text{yr}^{-1}$.
See the bottom panel of Fig.~\ref{fig:SubDiffLocal}. We also tried a soft upper boundary similar to the potential energy of a coiled spring; although it improved the pdf somewhat, it did not significantly improve the overall cost, so we do not include it here to reduce the number of free parameters. 

As the main result of our modeling efforts, the $g_1$ subdiffusive model yields an accurate approximation of the Mercury instability probability statistics for the $N$-body model both from 1--5~Gyr  and from 5--40~Gyr (Fig.~\ref{fig:SubDiffHits}).
The ability of the $g_1$ subdiffusive model to reproduce the Mercury instability statistics on timescales less than 10~Gyr is primarily due to larger variations in $g_1$ on timescales less than 0.3~Gyr and represents a significant advance beyond the previous $g_1$ diffusive model.

\subsection{Tuning Algorithm} \label{sec:tuning}

To fine-tune the values of $\alpha$, $r$, $g_1^{\downarrow}$, and $g_1^{\uparrow}$ in the subdiffusive model presented in section~\ref{sec:subdiffusive}, we introduce the cost function
\begin{equation*}
    C = 10 \cdot C_{\rm prob} + C_{\rm log} + C_{\rm disp} + 1000 \cdot C_{\rm hist},
\end{equation*}
where the components of the cost function are defined as follows:
\begin{itemize}
\item $C_{\rm prob}$ is the sum of differences in the instability probabilities $p$ for the subdiffusive model and the $N$-body model.
The sum is performed over all times $t = 0, 0.01, \ldots, 40\, \text{Gyr}$
\item $C_{\rm log}$ is the sum of differences in the weighted log instability probabilities $\log p / t$ for the subdiffusive model and the $N$-body model.
The sum is performed over all times $t = 0, 0.01, \ldots, 40\, \text{Gyr}$
for which $p > 0$ in the $N$-body model.
\item $C_{\rm disp}$ is the sum of differences in the weighted log mean square displacements $\log \langle |\Delta g_1|^2 \rangle / \Delta t$ for the subdiffusive model and the $N$-body model.
The sum is performed over all time lags $\Delta t = 0.01, \ldots, 5.0\, \text{Gyr}$.
\item $C_{\rm hist}$ is the sum of differences in the pdfs (histograms) of $g_1$ position for the subdiffusive model and the $N$-body model.
The sum is performed over the bins $[4.000\,''\,\text{yr}^{-1}, 4.025\,''\,\text{yr}^{-1}), [4.025\,''\,\text{yr}^{-1}, 4.050\,''\,\text{yr}^{-1}), \ldots$
\end{itemize}
In summary, the cost function penalizes disagreements in the long-time instability probabilities ($C_{\rm prob}$ component), the short-time instability probabilities ($C_{\rm log}$ component), the mean square displacements ($C_{\rm disp}$ component), and the pdfs of $g_1$ position ($C_{\rm hist}$ component).
These four components are scaled by the appropriate orders of magnitude so that they contribute similarly to the overall cost.

To minimize the overall cost $C$, we perform a grid search over possible parameter values and calculate probabilities based on $10^6$ realizations of the stochastic model, leading to the results listed in Table~\ref{tab:costs}. The table shows how the diffusive model of \citet{MogaveroLaskar2021} can be slightly improved by optimizing the parameters $r$, $g_1^{\downarrow}$, and $g_1^{\uparrow}$. In contrast, adopting the subdiffusive Hurst parameter $\alpha = 0.26$ dramatically reduces the cost. Compared to the original \citet{MogaveroLaskar2021} model, the best subdiffusive model has a $3 \times$ smaller cost. The subdiffusive model leads to the most improvement in $C_{\rm log}$, which is weighted toward earlier times, and $C_{\rm disp}$, as expected since it was motivated by the poor fit of the diffusive model to the mean square displacement. See Table~\ref{tab:params} for a list of the five converged parameter values for the subdiffusive model (four of which are free parameters).

\begin{table}[t]
\begin{center}
\begin{tabular}{ l|l|l|l } 
\textbf{parameters} & \textbf{ML21} & \textbf{diffusive} & \textbf{subdiffusive} \\
 \hline
 $10 \cdot C_{\rm prob}$ & $305$ & $310$ & $229$  \\ 
 $C_{\rm log}$ & $533$ & $363$ & $85$  \\ 
 $C_{\rm disp}$ & $610$ & $655$ & $63$ \\
 $1000 \cdot C_{\rm hist}$ & $446$ & $235$ & $219$  \\
 \hline
 $C$  & $1895$ & $1577$ & $595$ 
\end{tabular}
\end{center}
\caption{Cost function $C$ for different diffusive and subdiffusive models.
The ML21 model is taken from \citet{MogaveroLaskar2021}.
The ``diffusive'' and ``subdiffusive'' models fix the starting parameter $g_1^0 = 5.60\,''\,\text{yr}^{-1}$ and adapt the parameters $r$, $g_1^{\downarrow}$, and $g_1^{\uparrow}$ to minimize the cost.
The ``diffusive'' model fixes $\alpha = 0.50$, while the ``subdiffusive'' model adapts $\alpha$ to minimize the cost.}
\label{tab:costs}
\end{table}

%%%%%%%%%%%%%%%%%%%%%%%%%%%%%%%%%%%%%%%%%%%%%%%%%%%%%%%%%%%%%%%%%%%%
\section{Discussion} \label{sec:discussion}

The success of the $g_1$ subdiffusive model at approximating so many characteristics of the $N$-body model sets new challenges for the planetary dynamics community: First, why does $g_1$ subdiffuse rather than diffuse? Second, what is the physical cause of the restoring upper boundary on $g_1$? Given that the upper boundary prevents $g_1$ from resonating with $g_2\approx 7.45\,''\,$yr$^{-1}$, why isn't there a restoring lower boundary on $g_1$ that prevents it from reaching $g_5$ and thereby prevents Mercury instability events?

Clues to the answers to these questions may lie in the properties of a conservative, Hamiltonian system. The phase space of such systems are generically comprised of a mixture of regular and chaotic trajectories. The dynamics of two degree-of-freedom Hamiltonian systems are equivalent to those of area-preserving maps via the construction of  Poincarè return maps. The mixed phase space of two-dimensional area-preserving maps consists of elliptic periodic orbits surrounded by KAM curves that constitute ``islands'' embedded in a chaotic ``sea'' \citep[e.g.,][]{LichtenbergLieberman1992}. The KAM curves of these regular islands form strict barriers for trajectories in the chaotic regime. The ``stickiness'' of the borders of these regular islands could lead to behavior that can effectively be described as subdiffusion, similar to how diffusion on fractal materials can lead to subdiffusion \citep{henry2010introduction}.
In higher dimensions, the surviving KAM tori of regular trajectories would no longer impose strict topological constraints on the phase space accessible to chaotic orbits, but may still limit the range of excursions in a way that can be described by a soft, spring-like boundary. Of course this discussion is highly speculative, and more detailed research is needed to satisfactorily explain the dynamical properties of the Solar System we have identified in this paper.

\begin{figure}[ht!]
\centering
\includegraphics[width=\linewidth]{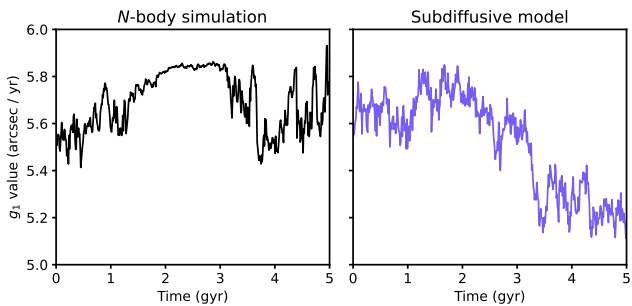}
\caption{Left: A sample trajectory of $g_1$ from the $N$-body model that demonstrates intermittency: sustained quiescent periods alternate with sustained active periods.
Right: A sample trajectory of $g_1$ from the subdiffusive model for comparison.}
\label{fig:g1_intermittency}
\end{figure}

The $g_1$ subdiffusive model is an improvement over the $g_1$ diffusive model, but it does not produce identical behavior to the $N$-body model. It is useful for improving understanding of the $N$-body model, not for replacing it. For example, the $N$-body trajectories show intermittency, transitioning from sustained quiescent periods to sustained active periods (Fig.~\ref{fig:g1_intermittency}). Periods of relative quiescence could be associated with proximity to islands of regularity. Finally, it is important to remember that an $N$-body code is not a perfect representation of reality either.

The subdiffusive model fits Mercury instability statistics from the $N$-body model much better than the diffusive model on timescales less than 5~Gyr, but we do not have enough $N$-body Mercury instability events on timescales less than 2~Gyr to thoroughly test the subdiffusive model on the shortest timescales. One option would be to generate large enough ensembles (possibly with $10^6$ members) using a high-order secular code \citep[e.g.,][]{MogaveroLaskar2021} to estimate the probability of a Mercury instability event on these short timescales. Alternatively, more $N$-body Mercury instability examples on shorter timescales could be obtained using Diffusion Monte Carlo \citep{Ragone2018heatwaves, webber2019practical,ragone2021rare,abbot2021rare} or action minimization \citep{E2004mam,plotkin2019maximizing,woillez2020instantons,schorlepp2023scalable} rare event schemes, aided by machine learning predictor functions \citep{ma2005nncommittor,chattopadhyay2020analog,Finkel2021learning,Miloshevich2022probabilistic,finkel2023revealing}.

%%%%%%%%%%%%%%%%%%%%%%%%%%%%%%%%%%%%%%%%%%%%%%%%%%%%%%%%%%%%%%%%%%%%
\section{Conclusions} \label{sec:conclusions}

The main conclusions of this paper are:

\begin{enumerate}
    \item The $g_1$ diffusive model significantly underpredicts the Mercury instability probability relative to an $N$-body model on timescales less than 5~Gyr, which is the physically relevant timescale for the future of the Solar System. The underprediction results from the fact that the $g_1$ diffusive model produces too small variations of $g_1$ on timescales less than $\sim$0.3~Gyr.
    \item We are able to fit $N$-body Mercury instability statistics on timescales of less than 5~Gyr as well as longer timescales using the $g_1$ subdiffusive model. We tune the model using 
    the mean square displacement of $g_1$, probability density function (pdf) of $g_1$ from $N$-body simulations, and the fraction of $N$-body simulations that reach the instability threshold.
\end{enumerate}

%%%%%%%%%%%%%%%%%%%%%%%%%%%%%%%%%%%%%%%%%%%%%%%%%%%%%%%%%%%%%%%%%%%%
%\begin{acknowledgments}
\vspace{1 cm}

We thank Dan Fabrycky for extensive feedback on an early draft of this paper. We thank an anonymous reviewer for excellent feedback. This work was completed with resources provided by the University of Chicago Research Computing Center. D.S.A acknowledges support from NASA grant No. 80NSSC21K1718, which is part of the Habitable Worlds program. 
R.J.W. was supported by the Office of Naval Research through BRC Award No. N00014-18-1-2363 and the National Science Foundation through FRG Award No. 1952777, under the aegis of Joel A. Tropp.  D.M.H acknowledges support from the CycloAstro project.
J.W. acknowledges support from National Science Foundation through award DMS-2054306 and from the Advanced Scientific Computing Research Program within the DOE Office of Science through award DE-SC0020427. D.S.A. and J.W. acknowledge support from the Army Research Office, grant number W911NF-22-2-0124.
%\end{acknowledgments}
\\
\\
\\
%\software{}
This research made use of the open-source projects Jupyter \citep{Kluyver2016jupyter}, iPython \citep{PER-GRA:2007}, and matplotlib \citep{Hunter:2007}.
%%%%%%%%%%%%%%%%%%%%%%%%%%%%%%%%%%%%%%%%%%%%%%%%%%%%%%%%%%%%%%%%%%%%
%\appendix

%%%%%%%%%%%%%%%%%%%%%%%%%%%%%%%%%%%%%%%%%%%%%%%%%%%%%%%%%%%%%%%%%%%%

\end{document}